\documentclass[11pt,twoside]{article}

\usepackage{asp2004}
\usepackage{epsf}
\usepackage{psfig}
\usepackage{lscape}

\begin{document}
\title{New Photometric Observations of \boldmath $\sigma$ Ori E}   %%% Fill in title
\author{Mary Oksala \& Rich Townsend}   %%% Fill in author names
\affil{Bartol Research Institute, University of Delaware, Newark, Delaware 19716, USA}    %%% Fill in author affiliations

\begin{abstract} %%% Abstract to run on from here.
We present new \emph{UBVRI} observations of the magnetic Bp star
$\sigma$ Ori E. The basic features of the star's lightcurve have not
changed since the previous monitoring by \citet{Hes1977}, indicating
that the star's magnetosphere has remained stable over the past three
decades. Interestingly, we find a rotation period that is slightly
longer than in the \citet{Hes1977} analysis, suggesting possible
spindown of the star.
\end{abstract}

%%% MAIN BODY OF TEXT GOES HERE. CONSULT "INSTRUCTIONS FOR AUTHORS USING
%%% LATEX2E MARKUP", SECTIONS 2.3-2.6 FOR HELP WITH EQUATIONS, FIGURES,
%%% AND TABLES.

%\section{}   %%% Top level section head (remove "%" symbol)
%\subsection{}   %%% Second level section head (remove "%" symbol)
%\subsubsection{}   %%% Lowest level section head (remove "%" symbol)
%\section*{}	%%% Unnumbered top level section head (remove "%" symbol)
%\subsection*{}   %%% Unnumbered second level section head (remove "%" symbol)

\section{Observations}

\begin{figure}[t!]
\plotone{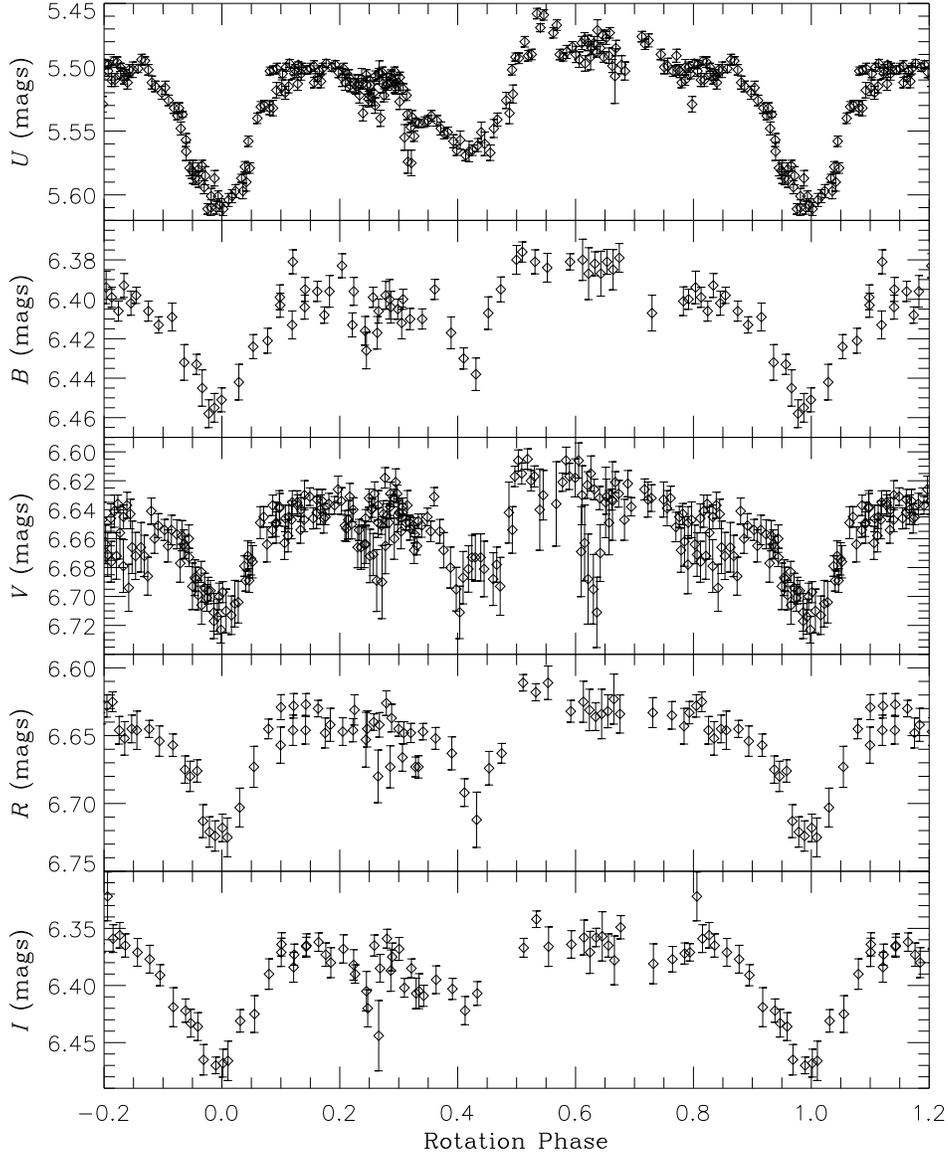}
\caption{Observed \emph{UBVRI} light curves, phased over an adopted
rotation period of 1.190858\,d.}
\label{fig:obs}
\end{figure}

The helium-strong star $\sigma$ Ori E (B2Vpe) is characterized by
1.19\,d-periodic modulations in its H$\alpha$ emission, helium
absorption-line strengths, visible and near-IR photometric indices, UV
continuum flux and resonance-line strengths, 6\,cm radio emission,
linear polarization, and longitudinal field strength. The most current
comprehensive photometric observations of the star are by
\citet{Hes1977}.  Recently, \citet[][hereafter T05]{Tow2005} have
demonstrated that the \emph{Rigidly Rotating Magnetosphere} (RRM)
model developed by \citet{TowOwo2005} can furnish a good fit to these
observations (see also Townsend, these proceedings).

In this contribution, we present fresh photometric observations of
$\sigma$ Ori E, obtained with the dual aims of further testing the RRM
model and searching for any changes in the star's circumstellar plasma
distribution. Beginning 17th November 2004, we conducted 14 nights'
Johnson-Cousins \emph{UBVRI} CCD photometry of the $\sigma$ Ori
system, using the 0.9 m CTIO telescope made available through the
University of Delaware's membership of the SMARTS consortium. The
filter sequence \emph{UVUVUVBRI} was adopted, and a neutral density
filter was employed to prevent CCD saturation. The choice of too high
a filter extinction on the first night rendered the data from that
night unusable.

The observations were reduced using standard procedures (flat
fielding, zero subtraction etc.) in \textsc{iraf}. PSF fitting was
employed to extract differential photometric indices for $\sigma$ Ori
E, which were then calibrated against nearby $\sigma$ Ori AB using the
colors published by \citet{Ech1979}. We do not employ a photometric
standard star, since we are primarily interested in \emph{changes} to
the brightness of $\sigma$ Ori E. The resulting light curves, phased
over the star's rotation period, are shown in Fig.~\ref{fig:obs}.

\section{Discussion}

\begin{figure}[t!]
\plotone{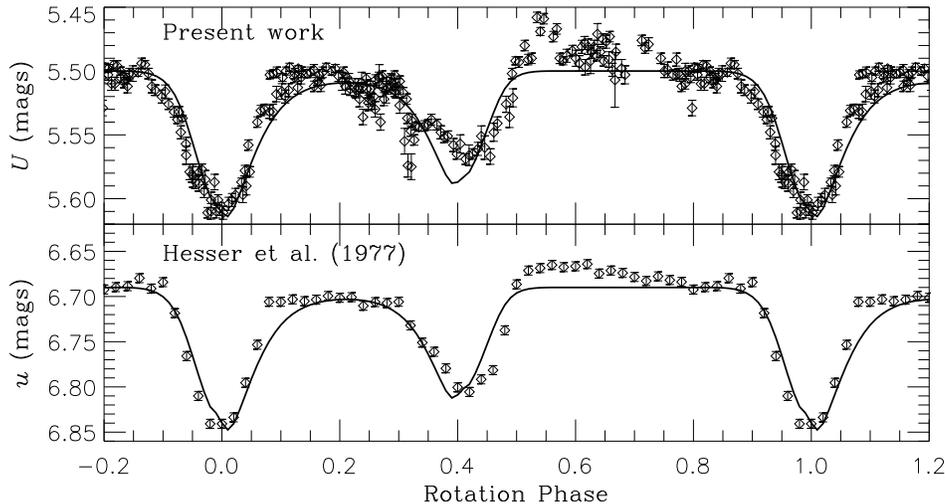}
\caption{Comparison between the Str\"{o}mgren \emph{u}-band light curve
obtained by\citet{Hes1977}, the Johnson \emph{U}-band light curve presented
here, and the RRM model for $\sigma$ Ori E devised by T05 (solid
line).}
\label{fig:mod}
\end{figure}

In Fig.~\ref{fig:mod}, we compare our \emph{U}-band observations with
both the \emph{u}-band light curve published by \citet{Hes1977} and
the predictions of the T05 RRM model. The primary minimum in both
light curves arises when dense magnetospheric plasma transits the
stellar disk; the depth and timing of this minimum in the RRM model is
adjusted to match the observations, and agreement between the two is
therefore guaranteed. The RRM model underestimates the star's
brightness at first maximum, when the magnetosphere is viewed near
edge-on. The deficit is most significant in the \emph{U}-band
observations, but can also be seen in the \emph{u}-band data. This
suggest that the star's inclination and/or magnetic obliquity may be
larger than the values ($i=75^{\circ}$, $\beta=55^{\circ}$) adopted by
T05.

The secondary light minimum is situated at the same phase $\sim 0.42$
in both observational datasets, indicating that the star's magnetic
field configuration has been stable over the past three
decades. However, the minimum is weaker in the \emph{U}-band curve,
suggestive of a change in the relative distribution of plasma within
the magnetosphere. The pseudo-emission feature in the \emph{u}-band
data at phase $\sim 0.55$, which the RRM model is unable to reproduce,
is still present in the \emph{U}-band curve. It seems likely that this
feature, which is also seen in the \emph{BVR} data
(Fig.~\ref{fig:obs}), corresponds to a photospheric inhomogeneity near
the magnetic pole.  We note that the period
$P=1.190858\pm0.000001\,{\rm d}$ required to achieve correct phasing
is slightly longer than that found by \citet{Hes1977}. Confirming
suggestions of period lengthening by \citet{Rei2000}, this may
represent evidence for the first case of magnetic spin down in an
early-type star (although see also Mikul\'{a}\v{s}ek, these
proceedings).

\acknowledgements %%% Text of acknowledgements runs on after this command.

This research has been partially supported by US NSF grant AST-0097983
and NASA grant LTSA04-0000-0060.

%%% THE BIBLIOGRAPHY
%%%
%%% CONSULT SECTION 3 OF "INSTRUCTIONS FOR AUTHORS" FOR HOW TO USE NATBIB.
%%% AUTHORS ARE ENCOURAGED TO USE EITHER THE "THEBIBLIOGRAPY" ENVIRONMENT
%%% BY UNCOMMENTING (DELETING THE "%" SYMBOL) THE COMMANDS BELOW, OR BY
%%% USING THE BIBTEX ENVIRONMENT. TO FIND OUT WHICH IS APPLICABLE TO YOUR
%%% CONTRIBUTION, CONSULT THE VOLUME EDITORS FOR YOUR PROCEEDINGS.
%%%

%\begin{thebibliography}{}
%\bibitem[]{}
%\bibitem[]{}
%\bibitem[]{}
%\bibitem[]{}
%\bibitem[]{}
%\bibitem[]{}
%\bibitem[]{}
%\bibitem[]{}
%\bibitem[]{}
%\bibitem[]{}
%\bibitem[]{}
%\bibitem[]{}
%\end{thebibliography}

\end{document}